%% file: template.tex
\documentclass{article}

\usepackage{arxiv}

\usepackage[utf8]{inputenc} 
\usepackage[T1]{fontenc}    
\usepackage{hyperref}       
\usepackage{url}            
\usepackage{booktabs}       
\usepackage{amsfonts}       
\usepackage{nicefrac}       
\usepackage{microtype}      
\usepackage{lipsum}
\usepackage{url}
\usepackage{graphicx}
\usepackage{subfig}

\title{Describing Alt-Right communities and their discourse on Twitter during the 2018 US mid-term elections}

\author{
  A\'ngel Panizo-LLedot\\
  Computer Science Department\\
  Universidad Aut\'{o}noma de Madrid, Spain\\
  \texttt{angel.panizo@uam.es} 
  \And
  Javier Torregrosa \\
  Biological and Health Psychology Department\\
  Universidad Aut\'{o}noma de Madrid, Spain\\
  \texttt{francisco.torregrosa@uam.es} 
  \And
  Gema Bello-Orgaz\\
  Departmento de Sistemas Inform\'{a}ticos\\
  Universidad Polit\'{e}cnica de Madrid, Spain\\
  \texttt{gema.borgaz@upm.es} 
  \And
  Joshua Thorburn\\
  RMIT University\\
  Melbourne, Victoria, Australia\\
  \texttt{joshthorburn18@gmail.com}
  \And
  David Camacho\\
  Departmento de Sistemas Inform\'{a}ticos\\
  Universidad Polit\'{e}cnica de Madrid, Spain\\
  \texttt{david.camacho@upm.es} 
}

\begin{document}
\maketitle

\begin{abstract}
The alt-right is a far-right movement that has uniquely developed on social media, before becoming prominent in the 2016 United States presidential elections. However, very little research exists about their discourse and organization online. This study aimed to analyze how a sample of alt-right supporters organized themselves in the week before and after the 2018 midterm elections in the US, along with which topics they most frequently discussed. Using community finding and topic extraction algorithms, results indicated that the sample commonly used racist language and anti-immigration themes, criticised mainstream media and advocated for alternative media sources, whilst also engaging in discussion of current news stories. A subsection of alt-right supporters were found to focus heavily on white supremacist themes. Furthermore, small groups of alt-right supporters discussed anime, technology and religion. These results supported previous results from studies investigating the discourse of alt-right supporters. 
\end{abstract}

\keywords{polarization, far-right, community finding, topic extraction, political extremism}

\input{introduction.tex}

\input{related-work.tex}
\input{methodology.tex}
\input{results.tex}
\input{discussions.tex}

\section*{Acknowledgements}
This work has been supported by several research grants: Spanish Ministry of Science and Education under  TIN2014-56494-C4-4-P grant (DeepBio) and Comunidad Aut\'onoma de Madrid under S2013/ICE-3095 grant (CYNAMON).

%
%
\bibliographystyle{unsrt}
\bibliography{template}

\end{document}

%% file: introduction.tex
\section{Introduction}

In recent years, far-right political parties and candidates have achieved increased electoral success across various Western states. Furthermore, terrorist attacks by far-right extremists, such as in Norway and New Zealand, demonstrate the risk that this extremism poses. Although it has a long history, the far-right has flourished in recent years, with the effective use of social media strongly contributing to this rise. While the alt-right itself is “highly decentralized”, with no official leaders, ideology or political party, this broad far-right amalgamation has effectively used social media to attract followers and influence political discourse. Indeed, the alt-right played a key supporting role in the election of US President Donald Trump. Alt-right supporters have used a variety of social media networks to spread content and engage in political discussion, including Twitter, Facebook, Reddit, 4Chan and YouTube. Notably, in response to social media companies increasingly removing hate speech and extreme far-right accounts, alt-right supporters have recently congregated on less-restrictive websites, such as Gab, an alternative to Twitter. 


While the discourse of other white extremist groups has been previously studied by academics \cite{graham2016inter}, there are few articles concerning how alt-right sympathizers communicate and interact online. However, the alt-right is a difficult group to study due to the movement’s fragmentation and lack of ideological agreement. With these limitations in mind, the present research aims to analyze how a set of alt-right supporters communicated on Twitter during the 2018 US midterm elections, a politically relevant period. To evaluate this, two questions were posed. First, \emph{"Are there different subsections of alt-right followers?"}, and second, \emph{"What are the topics discussed in these groups?"}.

%% file: related-work.tex
\section{Related work}



This section first describes the alt-right discourse and it similarities and dissimilarities with other ultra-conservative groups, before reviewing common techniques for doing online discourse analysis.

\subsection{Alt-Right discourse}

Firstly, opposition to immigration, particularly from the Middle East, Africa or Latin American countries, is a central political stance of alt-right supporters and the far-right. For example, the refugee crisis stemming from the Syrian Civil War led to far-right groups using hashtags to reject European countries accepting these immigrants, \cite{kreis2017refugeesnotwelcome}, such as \#Ausländerraus (“Foreigners out!”), \#EuropeforEuropeans or \#refugeesnotwelcome. Furthermore, alt-right followers have been supportive of Trump’s strict immigration policy proposals  \cite{greven2016rise}. Lyons  \cite{lyons2017ctrl} found that original manifestos from prominent ideological thinkers of the alt-right were deeply rooted in a rejection of non-white immigration. Similarly, alt-right supporters commonly use racist language to describe immigrants, much like other far right groups \cite{hawley2017making}. Indeed, the alt-right has popularised the use of white supremacist hashtags such as “\#ItsOkToBeWhite, or \#WhiteGenocide \cite{thorburn2018measuring}. Supporters of the alt-right frequently attack minorities and are especially hostile towards black people, Jews and Muslims. Consequently, while they are often aligned with Trump, alt-right followers have criticized him for being too soft against Saudi Arabia and Israel because they symbolically represent Islam and Judaism \cite{mirrlees2018alt}.

Criticism of mainstream media is also common in alt-right discourse \cite{lyons2017ctrl}, with this often mirroring President Trump’s frequent lamentations about “fake news”. A study conducted by Forscher \& Kteily \cite{forscher2017psychological} found that alt-right supporters tend to have high levels of suspicion towards mainstream media sources (i.e, New York Times, CNN, etc.) and strong trust in alternative media sources, such as the conspiratorial InfoWars. Conspiracy theories are also commonly believed alt-right followers. For example, it is commonly believed that certain media companies are actively trying to bring down the Trump presidency, or that a secretive Jewish elite hold enormous power \cite{finkelstein2018quantitative}. 

Alt-right supporters tend to be very politically active on various online social networks. Through the analysis of user descriptions from the online platform Gab (an online platform similar to Twitter) it was found that most of its far right users included references to Trump or his campaign slogans, conservative topics, religion or America \cite{zannettou2018gab}. Hashtags against Islam, about the alt-right itself were also found on the website. This analysis of Gab found that there posts by users contained 2.4 times more hate words than on Twitter, but less than half than those on 4chan. 4Chan itself, especially its /pol/ (‘Politically Incorrect’) message board, has been strongly linked with providing a fertile breeding ground for the alt-right to develop as a movement \cite{nagle2017kill}. 

Although many of the ideological underpinnings of the alt-right are shared with and stem from earlier and contemporary far-right movements, the alt-right has differed in some respects too. For example, although some alt-right supporters are fervently religious, they often differ strongly from the Evangelical conservative Christian right in the US. Indeed, “cuck”/”cuckservative”, the frequently used alt-right pejorative, derives from a sexual act. This slang is also representative of another characteristic of the alt-right: its ability to use humour to attract supporters. By using ironic, dark-humoured satire, the alt-right has succeeded in flippantly spreading white supremacist ideology on various online platforms \cite{hine2017kek}. Further, the alt-right has used internet memes, as well as creating its own imagery and slogans to attract supporters. 

\subsection{Techniques for online discourse analysis}




A number of technical approaches can be used to analyse online communications. Depending on the focus of the research (e.g. text, interactions, frequency of writing, etc.), certain methods are more appropriate. Natural Language Processing (NLP), which includes sentiment analysis, topic extraction or linguistic inquiry, is a common approach in analysing large textual datasets. Examples of this includes a study by Torregrosa et al \cite{torregrosa2019linguistic} where the tweets of jihadi supporters on Twitter was examined using Linguistic Inquiry Word Count (LIWC), or in a study by Tumasjan et al \cite{tumasjan2010predicting} where sentiment analysis was used to analyze tweets referring to political parties.

Understanding how people and groups interact on Online Social Networks (OSNs) is important when analyzing discourses. Social Network Analysis (SNA) techniques are especially useful to study these interactions. While some SNAs are focused on extracting common characteristics of a network, such as the density or its diameter, others focus on characterizing properties of the users of the network. For example, analyzing the Homophily of the network or finding cohesive groups of users (communities) can be an effective approach. ‘Community finding’ is one of the most commonly used techniques to investigate and analyze OSNs.

Several studies have combined community finding with NLP techniques. Among them, three main approaches can be found in the literature. The most frequent approach combines NLP techniques and community finding into a singular process. In this approach, communities are not only consistent in a structural manner, but also in the topics they discuss. Feller et al. \cite{feller2011divided} used this type of method when applying a network analysis to contextualize the users political preferences during the 2016 Germany elections. The second approach apply both process simultaneously but separated, comparing the outcomes later. An example of this approach can be found in Surian et al \cite{surian2016characterizing}, where they analyze the discussions about HPV vaccines on Twitter. Finally, the third approach initially applies the community finding process, and then applies NLP techniques in each community independently. This was used by Yin et al \cite{yin2012latent} in a study comparing two datasets, with one containing computer science journals and the other composed of tweets that included the keywords "Obama" or "media". In the present research, this third approach was applied to examine the tweets and the communities.

%% file: methodology.tex
\section{Methodology}

\subsection{Data description}

The dataset used in this study contained 52903 tweets published by 123 alt-right Twitter accounts collected between the $30^{th}$ of October and the $13^{th}$ of November, 2018. This period corresponds to the weeks before and after the 2018 US midterm elections. In addition to the text of each tweet, information such as publication date, retweets or hashtags was gathered. The selected alt-right Twitter accounts were chosen from a dataset created by Thorburn, Torregrosa and Panizo \cite{thorburn2018measuring}.  The alt-right users published between 2 and 2590 tweets during the data collection period. In fact, $\approx60\%$ of the accounts tweeted every day of this period and $\approx80\%$ of the users in the dataset tweeted on at least 10 different days. 

\begin{figure*}
\centering
\begin{tabular}{c c c}
\subfloat[] {\includegraphics[width=0.31\linewidth]{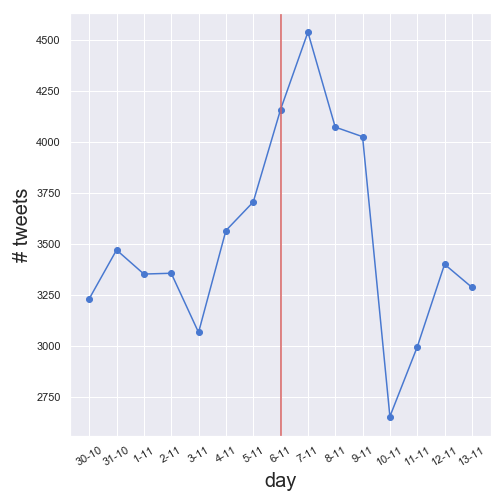}}
&
\subfloat[] {\includegraphics[width=0.31\linewidth]{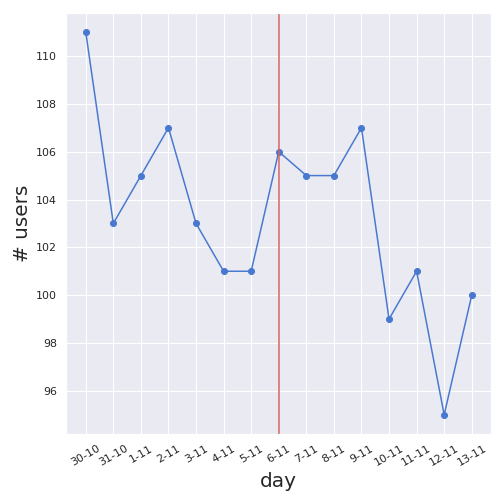} }
&
\subfloat[] {\includegraphics[width=0.31\linewidth]{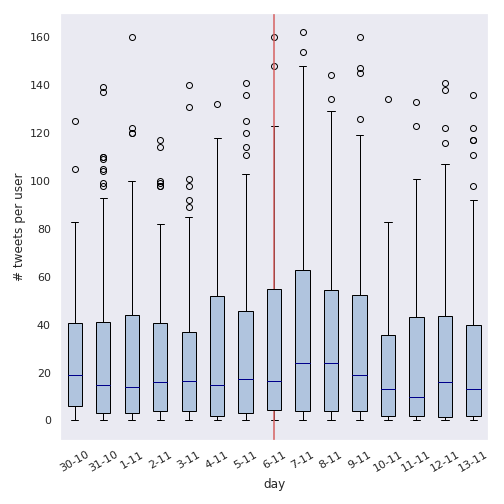} }
\\
\end{tabular}
\caption{Subfigure (a) shows the number of tweets published each day. The x-axis contains the days and the y-axis the number of tweets published. Subfigure (b) shows the number of accounts that published something every day of the data collection period. The x-axis contains the days and the y-axis the number of accounts that tweeted on each day. Subfigure (c) shows the distribution of the number of tweets published by an account for each day. The x-axis contains the days and the y-axis the number of tweets published. The red line marks election day (November 6).}

\label{fig:number-tweets}
\end{figure*}

As shown in Figure \ref{fig:number-tweets} (a), the number of tweets published per day was between 2656 and 4539 tweets, with increased activity surrounding election day (November 6, marked as a red vertical line), with this peaking on November 7, the day immediately after the elections. However, in regards to the number of users posting each day (shown in Figure \ref{fig:number-tweets} (b)), it can be noticed that the same spike in activity is not present. Finally, regarding the daily distribution of tweets published by each account (see Figure \ref{fig:number-tweets} (c)), the median and the whiskers of the box blot diagram indicate that this also increased surrounding election day.


\subsection{Protocol}

Once the data was gathered to create the dataset, this information was divided and analyzed using community finding and topic detection algorithms. The phases of this process is detailed as follows:

\begin{enumerate}
    \item \textbf{Data Extraction}: The data collected to perform the analysis of the alt-right discourse was extracted from the Twitter API. Every tweet published by the users in the selected two time periods was included for analysis. The extracted tweets are stored in a database (MongoDB) together with their related information, such as creation date, location, source, retweet count, etc.

     \item \textbf{Data Preprocessing}: Two re-tweet networks of users were created, representing the interactions between all users before and after the election day. To create each network, users were considered as the network nodes, and their relationships represented the edges. The relationships were established using the re-tweets. 
     
     \item \textbf{Community Detection of Alt-Right Users}: The detection of relevant groups of users within the OSN can be addressed by the application of \textit{community detection algorithms} in the re-tweet networks. Before applying the community search algorithm, a filter was used to get the largest connected component of each re-tweet network. Then, the most similar users were grouped together using the Clauset-Newman-Moore greedy modularity maximization method \cite{clauset2004finding,Newman2006}.  
     
     \item \textbf{Topic Extraction by the Communities}: Once the different communities were obtained for each re-tweet network, a topic modeling technique was applied to identify the topics covered by a collection of texts. With this aim, the Latent Dirichlet Allocation (LDA) model \cite{blei2003latent} was fitted and applied for each community, using only the text of the tweets published by the users belonging to the particular community. In addition, the most mentioned hashtags for each community were extracted.
     
     \item \textbf{Analysis of the Social Discourse and Evolution of the Communities}: The analysis of the discourse of each alt-right community on Twitter was conducted using the evolution of its relevant topics. For this purpose, topics were manually categorized according to the general themes that were found to be relevant in the discourse of the alt-right groups in the related work section. A summary of the categories can be found in Table 1.

\end{enumerate}

\begin{table*}
\centering
\begin{tabular}{|l|p{5cm}|p{3,5cm}|l|}
\hline
\textbf{Categories} & \textbf{Example keywords}                                                                                              & \textbf{Example hashtags}                               & \textbf{Code} \\ \hline
Racial discourse   & White, Racist, Black, Hate, Discrimination, Diversity, Race, Israel, Semitic, Jewish, Jerusalem, Islam, Iran, Pakistan & \#Iran, \#WhiteGenocide, \#ItsOkToBeWhite               & RD            \\ \hline
Politics            & Trump, Obama, Hillary, President, Right, Left, Party, Republican, Democrats, Maga, Antifa                              & \#Maga, \#Democrats, \#AmericaFirst                     & P             \\ \hline
Immigration         & Border, Immigration, Caravan, Migrants, Citizenship, Wall, Asylum, Wave, Illegal, Invasion, Refugees                   & \#BirthrightCitizenship, \#WalkAway, \#StandUpForEurope & I             \\ \hline
Media               & Media, News, Fake, Gab, Alternative media, Press, Hivemind, Propagandistic, CNN, Conspiracy, Journalist                & \#Killstream, \#WSJKillsKids, \#CNN                     & M             \\ \hline
Elections           & Election, Vote, Campaign, Senate, Mid-term, Rally, Ballots                                                             & \#Midterms2018, \#VoteRed, \#ElectionDay                & E             \\ \hline
Policy debate       & Money, Taxes, Taxation, Marxism, Socialism, Communist, Freedom, Nationalism, Brexit                           & \#TaxationIsTheft                                       & PD            \\ \hline
\end{tabular}

\caption{Categories used to classify the topics covered in the alt-right communities.}

\label{tab:Categories}
\end{table*}

%% file: results.tex
\section{Results}


Table \ref{tab:Prev} summarizes the communities extracted from the sample before the election, along with information relating to the topic extraction process, the most mentioned hashtags and the categories in which the topics were included. Nearly all the communities contained politically focused discourse, including references to the elections (\#ElectionDay), right-wing political slogans ("Make America Great Again", \#MAGA) or other relevant political groups (\#Anonymous).

Most of the groups include words associated with far-right ideologies and the alt-right more specifically. Racist discourse was identified in many of the communities. This can be subcategorized into white supremacy (communities 1, 3, 4, 10 and 12), antisemitism (communities 1 and 10) and Islamophobia (community 10). Group 12 was found to be especially extreme in its language, using hashtags such as \#WhiteGenocide and \#WhiteLivesMatter, which are associated with the alt-right and white supremacism. 

The references to the media were divided between pro-comments (comments defending alternative media) and anti-comments (comments against traditional media). In the first subcategory, there are references to media related with far-right movements (\#Killstream, \#GAB, \#RedNationRising). There are also references to Gab (some users presented a link to their Gab profile on their Twitter descriptions). The second subcategory includes criticisms against traditional media, with this including hashtag campaigns against CNN ("\#CNNCredibilityCrisis) and the Wall Street Journal (\#WSJKillsKids). Criticism of traditional media outlets was especially notable in community 14, where CNN was mentioned several times.

As expected, immigrations was frequently mentioned, and this can be divided between references to illegal immigration from Mexico (e.g. communities 8 and 11) or the refugee crisis in Europe (e.g. communities 5, 10 and 11). The words "Trump" and "Border" appear together in some groups (e.g. community 8), which indicates that users commonly discussed the Trump administration’s immigration policies.

Interestingly, three groups seem to talk about isolated topics. Community 6 is the only one that focused on Brexit, taxes and political ideologies (Communism, Socialism, Neoliberalism, etc.). Furthermore, communities 9 and 15 centred around discussion of anime and religion respectively. These outliers illustrate some of the distinctions and demographics that the alt-right is composed of, which is elaborated on further in the discussion section. 

Concerning the week after the election, Table \ref{tab:compostresults} shows changes in some of the communities. Firstly, there was an increase in tweets about the election itself (communities 1, 2, 3, 8, 9, 12 and 13), with the word “election” and the \#ElectionDay being used more. Notably, \#Broward featured prominently, with this hashtag referring to difficulties in vote counting in the county of Broward, Florida. Various stories from right-wing media and political figures alleged that there had been a conspiracy to rig the election in favour of Democrats in this county. This alleged conspiracy was especially prominent in communities 2, 3 and 12, where the \#StopTheSteal was frequently used. 

\begin{figure*}
\centering
\noindent
\makebox[\textwidth][c]{\includegraphics[width=1.25\linewidth]{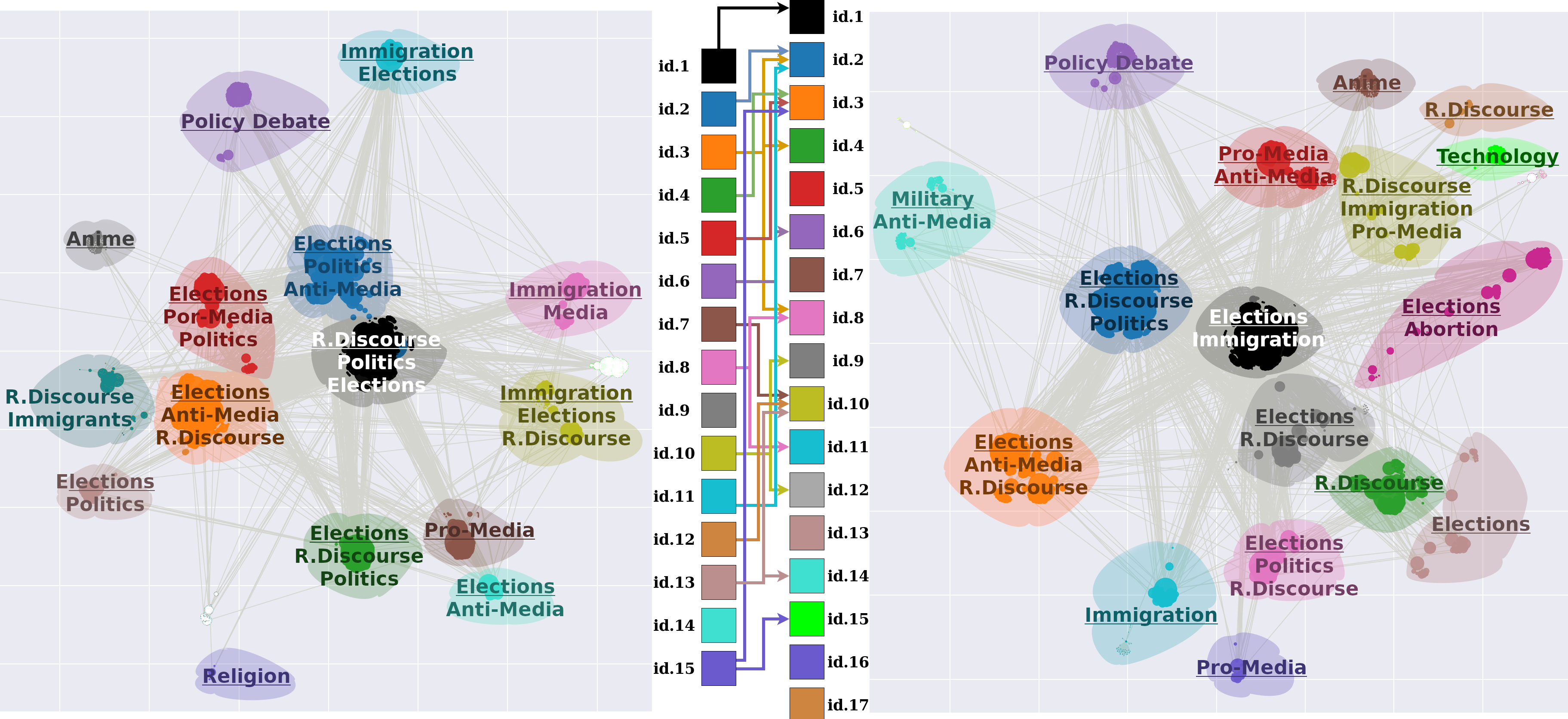}}
\caption{ On the left, communities detected before the election day. On the right, communities detected after the election day. Each community is labeled with their topics category, the underscored text represents main topic categories. An arrow joining two community on the legend indicates the transference of a community between snapshots.}
\label{fig:comms}
\end{figure*}

While no community maintained their original structure from the previous week, some of them conserved their topics and their most relevant users, as seen on Fig. \ref{fig:comms}. This was the case in community 6, which featured political debate in both weeks; community 7, which was dedicated to discussing anime (like community 9 from previous week); and community 10, which was the most extreme community in the second week too, using exactly the same hashtags that community 12 had used previously.  

Most of the topics identified in the week before the election were also found in the second week. White supremacist (communities 2, 3, 8 or 10) and antisemitic (communities 9 and 17) discourse was maintained, but there was no community that included a clear reference to Islam in the two weeks that our data covers. Still, discussion of immigration featured in communities 1, 10 and 11. Again, there were comments against traditional media (communities 3, 5, 14 and 16), including with references to "fake news" (Community 16). There were also tweets supporting alternative media (Community 5, 10 and 16). Interestingly, Gab and the hashtag \#KillStream (this hashtag is for a far-right podcast of the same name) are mentioned again. Lastly, several communities mention politics in their discussions (e.g. Community 8).

Finally, there were three new categories in the week following the election. First, community 14 contained references to war and military forces, mentioning the hashtags \#Veteran and \#Iran. The second was community 15, which was focused on debates about technology in general (including references to blockchain, crypto currency, AI, etc.). Lastly, debate about abortion was mentioned by community 12. 

\begin{table}
\begin{center}
\addtolength{\leftskip} {-2cm} 
\addtolength{\rightskip}{-2cm}
\scalebox{0.93}{
		\begin{tabular}{|c|c|p{7.5cm}|c|c|}
		\hline
		\textit{\textbf{Id.}} & \textit{\textbf{N.U.}} & \textit{ \textbf{Top 3 Topics}} & \textit{\textbf{Top 3 Hashtags}} & \textit{\textbf{Labels}} \\	
		\hline  
		  1 & 909 & people white new want america & Killstream & \textbf{RD(we, as)} \\ 
		    &  & got racist better hate nyt & MAGA & P \\ 
		    &  & anti gop twitter election black semitic & WSJKillsKids & E \\ 
		\hline
		  2 & 745 & trump president obama media campaign & MAGA & \textbf{E} \\ 
		    &  & trump left border caravan people & Midterms2018 & P \\ 
		    &  & democrat gillum breaking party florida & VoteRed & M (anti) \\ 
		\hline
		  3 & 531 & really trump true girl love & WSJKillsKids & \textbf{E} \\ 
		    &  & good man bad today voting  & Killstream & M (anti) \\ 
		    &  & women gt hate american twitter & HappyHalloween & RD (w,b) \\ 
		\hline
		  4 & 294 &  vote democrats red voting trump & MAGA & \textbf{E} \\ 
		    &  & campaign wow maga know migrants  & BLEXIT & RD (we) \\ 
		    &  & trump president obama democrats america  & VoteRed & P \\ 
		\hline
		  5 & 274 &  vote trump president america florida & Iran & \textbf{E} \\ 
		    &  & democrats family brownley senate women california   & RedNationRising & M (pro) \\ 
		    &  & shawn company insurance stopped program   & MAGA & I \\ 
		\hline
		  6 & 205 &  good time woman high child & TaxationIsTheft & \textbf{PD} \\ 
		    &  & read sense world better vegans    & DonLemon &  \\ 
		    &  & people socialism hate free marxist   & mises & \\ 
		\hline
		  7 & 177 &  people online love media god  & VerifiedHate & \textbf{M (pro)} \\ 
		    &  & gab speech free hate left    & Update &  \\ 
		    &  & world getting think read gt & Gab & \\ 
		\hline
		  8 & 160 &  just think know trump middle  & EnemyOfThePeople & \textbf{I} \\ 
		    &  & trump media job usa hit   & wages &  M \\ 
		    &  & immigration wall country border asylum & manufacturing & \\ 
		\hline
		  9 & 119 &  starting rain evidently mustang sins   & jojo\_anime & \textbf{Anime} \\ 
		    &  & halloween anime game minmod mini  & azurlane &   \\ 
		    &  & eat bad hey bro crunch  & NoTatsukiNoTanoshii & \\ 
		\hline
		  10 & 115 &  trump obama white caravan migrant  & TheTruthCommunity & \textbf{I} \\ 
		    &  & trump president caravan migrant midterm  & FaroeIslands & E  \\ 
		    &  & democrats iran midterms sanctions kavanaugh  & Anonymous & RD(ai)\\ 
    	\hline
		  11 & 99 &  stop country home triangle lemon & Halloween2018 & \textbf{I} \\ 
		    &  & trump illegal beto cnn caravan texas   & HappyHalloween2018 & E  \\ 
		    &  & border invasion campaign southern illegal  & MAGA & \\ 
    	\hline
		  12 & 67 &  white hatred sick wshis girls & WhiteGenocide & \textbf{RD (we)} \\ 
		    &  & diversity itsokaytobewhite defund whitepeople word code  & StandUpForEurope & I  \\ 
		    &  & migrants refugees europe eu border & WhiteLivesMatter & \\ 
    	\hline
		  13 & 64 &   memes hillary tweet trump black & MAGA2018 & \textbf{E} \\ 
		    &  & timeline country economy train book  & BTFOMobRule & P  \\ 
		    &  & voteredtosaveamerica hivemind critical marxist theory & memewarmidterms & \\ 
    	\hline
		  14 & 63 &  trump campaign candidate gillum navy  & CNNCredibilityCrisis & \textbf{E} \\ 
		    &  & rofl voting sundaymorning mondaymotivation thewalkingdead  & ElectionDay & M (anti)  \\ 
		    &  & cnn vote obama gillum rally  & ElectionDay2018   & \\ 
    	\hline
		  15 & 4 &  american stones demons boy bad church & CavalierNationalism & \textbf{Religion} \\ 
		    &  & gt christ right blessed strates   & WhiteHonorKilllings &  \\ 
		    &  & protestant fascism authoritarian mary collectivist   &    & \\ 
		\hline		
		\end{tabular}
}
\end{center}
\begin{scriptsize}
\caption{Communities detected before the Election Day and the general topics on which their users have posted messages.}
\label{tab:Prev}
\end{scriptsize}
\end{table}

\begin{table}
\begin{center}
\addtolength{\leftskip} {-2cm} 
\addtolength{\rightskip}{-2cm}
\scalebox{0.9}{
		\begin{tabular}{|c|c|p{7cm}|c|c|}
		\hline
		\textit{\textbf{Id.}} & \textit{\textbf{N.U.}} & \textit{ \textbf{Top 3 Topics}} & \textit{\textbf{Top 3 Hashtags}} & \textit{\textbf{Labels}} \\	
		\hline  
		  1 & 784 & election county tucker broward antifa & MAGA & \textbf{E} \\ 
		    &  & white america black live press & Editorial & I \\ 
		    &  & democrat gillum breaking party florida & Broward &  \\ 
		\hline
		  2 & 758 & county ballots election votes florida & StopTheSteal & \textbf{E} \\ 
		    &  & trump acosta white states jim & ElectionNight & RD (we) \\ 
		    &  & year boom years women stop & Broward &  P \\ 
		\hline
		  3 & 571 & trump president just vote got & StopTheSteal & \textbf{E} \\ 
		    &  & broward ballots county florida fraud snipes & BrowardCounty & M (anti) \\ 
		    &  & white people election democrats steal & MAGA & RD (w,b) \\ 
		\hline
		  4 & 256 &  time day men fam long & WoolseyFire & \textbf{ RD (we)} \\ 
		    &  & people good don like white  & electionnight & \\ 
		    &  & right new fuck bad community  & ff &  \\ 
		\hline
		  5 & 215 &  streaming company home muslim think & Killstream & \textbf{M (pro,anti)} \\ 
		    &  & white pride thread beagle deadly    & WSJKillsKids &  \\ 
		    &  & sargon really youtube just people   & BREAKING &  \\ 
		\hline
		  6 & 178 & socialism marxist nationalism party british & MeToo & \textbf{PD} \\ 
		    &  & men group make things word  & AsiaBibi &  \\ 
		    &  & western christian asiabibi asylum british  & JimAcosta & \\ 
		\hline
		  7 & 157 & wow evangelion taught looks whoah & DELTARUNE & \textbf{Anime} \\ 
		    &  & real needs energy avi world &  &  \\ 
		    &  & time undertale aha know kinda  &  & \\ 
		\hline
		  8 & 149 &  white women vote yes god & ElectionDay & \textbf{E} \\ 
		    &  & trump election republican ballots vote  & Florida &  P \\ 
		    &  & broward county tucker twitter scott & Broward & RD (we) \\ 
		\hline
		  9 & 144 &  followers help million raise communist  & StopTheSteal & \textbf{E} \\ 
		    &  & watch just media trump kid & TaxationIsTheft &  RD (as) \\ 
		    &  & nationalism broward  hitler ballots county  & Nationalism & \\ 
		\hline
		  10 & 104 & white people problem asylum facebook migrants & WhiteGenocide & \textbf{RD (we)} \\ 
		    &  & defund discrimination whites sense & StandUpForEurope & I  \\ 
		    &  & gab migrant media free right facebook & WhiteLivesMatter & M(pro)\\ 
    	\hline
		  11 & 100 & border entry ports story italy  & 2A & \textbf{I} \\ 
		    &  & white fact german ohio language & FoxNews &   \\ 
		    &  & trump european imagine language jewish & France & \\ 
    	\hline
		  12 & 96 &  trump stop american democrat abortion & StopTheSteal & \textbf{E} \\ 
		    &  &  macron women michelle football saudi  & USMC & Abortion  \\ 
		    &  & florida election county broward fraud & MarineCorpsBirthday & \\ 
    	\hline
		  13 & 95 & sounds fuck shit cultural florida & VeteransDay2018 & \textbf{E} \\ 
		    &  & trump great non cruz stop  & Nature &   \\ 
		    &  & got wonder family black white & MAHw1d5 & \\ 
    	\hline
		  14 & 85 &  marine love saved children television & Iran & \textbf{Military} \\ 
		    &  & god bless noticed lady today  & Veteran & M (anti)  \\ 
		    &  & press jim acosta stopthesteal trump & MAGA   & \\ 
    	\hline
		  15 & 63 & retired proud crypto jeff major  & Blockchain & \textbf{Technology} \\ 
		    &  & president ai stage crypto blockchain & AI &  \\ 
		    &  & tonight broward skating sponsorship paccoin   & WashingtonElite   & \\ 
    	\hline
		  16 & 28 & news talking fake weird fbpe  & TwoMinuteSilence & \textbf{M (pro)} \\ 
		    &  & british little ffs great sure  & RemembranceDay2018 &  \\ 
		    &  & just white good make ve video & ArmisticeDay100   & \\ 
    	\hline
		  17 & 10 &  antifa pay polish fuck jewish  & MeToo & \textbf{RD (as)} \\ 
		    &  & wanking media deficit social migrant  & PolishIndependenceDayMarch &  \\ 
		    &  & trump voters punish jobs black jerusalem & SouthAfrica   & \\

		\hline		
		\end{tabular}
}
\end{center}
\begin{scriptsize}
\caption{Communities detected after the Election Day and the general topics on which their users have posted messages.}
\label{tab:compostresults}
\end{scriptsize}
\end{table}

%% file: discussions.tex
\section{Discussion and Future work}

This study aimed to analyze the communications of a group of alt-right supporters in the context of an important political event, in this case, the 2018 US midterm elections. Firstly, we demonstrated that there was increased Twitter activity in the days immediately surrounding the election. The outcomes of this study also demonstrates that sub-communities from this database were mostly unstable over the data collection period, but that many topics did maintain over this period. 

Regarding the communities, it was found that none maintained the same structure of users over both weeks. Some of the relevant users remained in the same communities from one week to another (table 2), but they did not represent more than the 20-40\% on average of the previous relevant users. Therefore, it was easier to interpret communities using the topic they discussed, than by the users that were included. If we compare the topics covered by the communities and their categorization, we can find that some of the communities did maintain over time. A perfect example would be the community using white supremacist hashtags (community 12 pre-elections, 10 post-elections), or community 6, whcih discussed the economy and general politics in both weeks. Therefore, while communities of users did change during the data collection period, similar topics were discussed in many of these groups. 

Indeed, the topics that our data identifies were commonly discussed reflected earlier research on the alt-right. For example, anti-immigrant sentiment \cite{greven2016rise,kreis2017refugeesnotwelcome,lyons2017ctrl} and racially derogatory comments \cite{hawley2017making} were common. Furthermore, hashtags such as \#WhiteGenocide and \#WhiteLivesMatter were used frequently in some communities, and these hashtags are strongly associated with white supremacy. As to be expected, discussion of current political news stories was frequent, with this frequently mirroring right-wing media talking points. Similarly, criticism of traditional news media (such as CNN and The Wall Street Journal) was detected, with this often mirroring President Trump’s denunciations of the press. In contrast, "alternative" media sources were frequently shared or discussed, such as from Kill Stream and Red Nation Rising, with this reflecting other research on the alt-right \cite{lyons2017ctrl}.

Interestingly, our results indicate that anime and technology were discussed by many alt-right supporters. These findings provide support to other literature which has linked internet subcultures to the growth of the alt-right. For example, in Kill All Normies, Angela Nagle navigates the link between \#GamerGate, 4Chan and even sections of anime fans with the rise of the alt-right. Similarly, in Mike Wendling’s book on the alt-right, he agrees with this assessment and also connects proponents of technologies like cryptocurrencies with the movement. The results of this study provides further evidence of this link between the alt-right and certain sections of anime fans and cryptocurrency proponents. 

The results of this study also indicate that there is a small section of alt-right supporters who commonly discuss religion. While alt-right supporters are typically hostile towards Judaism and Islam, the movement cannot be broadly characterised as Christian, as a large contingent appear to be irreligious \cite{hawley2017making}. Still, results here identified a small community of users where religion was discussed before the election and another where abortion was a central topic after the midterm elections. This is perhaps evidence of the small, paleoconservative element which is aligned with the alt-right, where opposition to abortion is a strongly held belief \cite{hawley2017making}.

Considering the similarity on the topics, the increased relevance of elections on the second week and the increasing on the tweeting rate until the day of the elections, we can suggest that the Alt-Right communicational activity grows when a political event is near. However, a study picking longer dates or making a deeper assessment on the daily evolution of the communities could help to confirm this hypothesis. Also, it could be useful to determine why communities are quite different, how and when do they break into different groups, and why the topics covered by this communities are maintained.

Overall, this research demonstrated that alt-right supporters were more active in the days immediately surrounding the 2018 US midterm elections. Furthermore, this study identified a number of sub-communities within the broader alt-right, that gathered around certain topics. Similar to previous literature on the alt-right, anti-immigrant sentiment, racism towards Jews and Muslims, as well as support for white supremacy, were all frequently reflected in the results. Also supporting earlier literature, our results identified smaller, yet still notable segments of the alt-right sample used in this study discussing anime, technology and religion.